\documentclass[preprint,prd,showpacs]{revtex4}
\usepackage{graphicx}
\usepackage{amsmath}
\usepackage[pdftex,draft=false,letterpaper=true,colorlinks=true,urlcolor=black,linkcolor=black,citecolor=black,bookmarksopen=false,bookmarksnumbered=true,breaklinks=true,pdfauthor={Antonio Mondragon},pdfcreator={Antonio Mondragon},pdftitle={Alternative Derivation of the Relativistic Contribution to Perihelic Precession},pdfkeywords={Schwarzschild,perihelion,perihelic,perihelia,precession,perihelial,Keplerian,Kepler,relativity,spacetime,Einstein,eccentric,eccentricity,potential,effective,relativistic,orbit,ellipse,elliptical,Mercury,planet,circular,Newton,discrepancy,correction,alternative,Newtonian,derivation,general,metric,approximate,coordinate,transformation,correspondence,principle,perturbative,reduced,radius,undergraduate,graduate,calculation,line,element,timelike,geodesic,variational,lagrange,invariant,angular,momentum,radial,equation,nonlinear,differential,conic,sections,spherically,symmetric,sun,solar,system,mars,observation,rate,momenta,semi-major,speed,light,angle,central,mass,matter,domain,validity,linearization,linearisation,physical,mondragon,lemmon}]{hyperref}
\usepackage[all]{hypcap}

\def\rs{r_\text{\!S}^{}}
\def\rmd{{\rm d}}

\begin{document}
\title{Alternative derivation of the relativistic contribution to perihelic precession}

%\date{\today}

\author{Tyler J. Lemmon}

\author{Antonio R. Mondragon}

\email{antonio.mondragon@coloradocollege.edu}

\affiliation{Department of Physics, Colorado College, Colorado Springs, Colorado 80903}

\begin{abstract}
An alternative derivation of the first-order relativistic contribution to perihelic precession is presented. Orbital motion in the Schwarzschild geometry is considered in the Keplerian limit, and the orbit equation is derived for approximately elliptical motion. The method of solution makes use of coordinate transformations and the correspondence principle, rather than the standard perturbative approach. The form of the resulting orbit equation is similar to that derived from Newtonian mechanics and includes first-order corrections to Kepler's orbits due to general relativity. The associated relativistic contribution to perihelic precession agrees with established first-order results. The reduced radius for the circular orbit is in agreement to first-order with that calculated from the Schwarzschild effective potential. The method of solution is understandable by undergraduate students.
\end{abstract}

\pacs{04.25.Nx}
\maketitle

\section{Introduction}\label{Sec_Intro}

The derivation of the relativistic contribution to perihelic precession is found in almost every general relativity textbook \cite{MTW,weinberg,rindler,PleKra,OR,dinverno,hartle,carroll,HobEfsLas,wald}. The standard approach is perturbative and results in a first-order prediction for the relativistic contribution to perihelic precession which is in agreement with observations. In particular, the agreement of this contribution with the observed shift of the perihelion for Mercury is one of the most celebrated verifications of Einstein's general theory of relativity \cite{einstein0,einstein1,einstein2,einstein3}, and it is important that students are presented with a convincing calculation of this result. We present an alternative approach to this problem, and obtain an approximate relativistic orbit equation. Perihelic precession arises as one of several small corrections to Kepler's orbits, and agrees with established first-order calculations. 

The path of a small test mass near a spherically symmetric central mass $M$ is uniquely described by the Schwarzschild line element \cite{MTW,weinberg,rindler,PleKra,OR,dinverno,hartle,carroll,HobEfsLas,wald,ovanesyan,schild,wang,schwarzschild,droste}:
\begin{equation}
\rmd s^2 = ( 1 - \rs/r) c^2\rmd t^2 - (1 - \rs/r )^{-1}\rmd r^2 - r^2 \rmd\varOmega^2, \label{SchwarzschildMetric}
\end{equation}
where $\rmd\varOmega^2=\rmd\theta^2 + \sin^2{\theta}\,\rmd\varphi^2$ is an infinitesimal element of solid angle, and $\rs \equiv 2GM/c^2$ is the Schwarzschild radius. ($G$ is Newton's universal gravitational constant, and $c$ is the speed of light in vacuum.) We parameterize timelike geodesics as $\rmd s^2=c^2\rmd\tau^2$, where $\tau$ is the proper time along the path of a test particle. The variational principle for a geodesic results in Lagrange's equations,
\begin{equation}
\frac{\rmd}{\rmd\tau} \frac{\partial \mathcal{L}}{\partial\dot{x}^\mu} - \frac{\partial \mathcal{L}}{\partial x^\mu} = 0,
\end{equation}
for each of $\{ x^\mu \} = \{ t, r, \theta, \varphi \}$, where $\dot{x}^\mu\equiv\rmd x^\mu/\rmd\tau$, and
\begin{equation}
\mathcal{L} = ( 1 - \rs/r ) c^2 \dot{t}^2 - ( 1 - \rs/r )^{-1} \dot{r}^2 - r^2 (\dot{\theta}^2 + \sin^2{\theta} \dot{\varphi}^2).
\end{equation}

Consider orbits in the plane defined by $\theta = \pi/2$, so that $\dot{\theta} = 0$. The invariants of the motion are given by Lagrange's equations for $\varphi$ and $t$:
\begin{equation}
-2\dfrac{\rmd}{\rmd \tau} [r^2 \dot{\varphi}] = 0,
\end{equation}
which implies that $\ell \equiv r^2\dot\varphi = \mbox{constant}$; and
\begin{equation}
2 c^2 \dfrac{\rmd}{\rmd \tau} [ (1 - \rs/r) \dot{t} ] = 0,
\end{equation}
which implies that $k \equiv (1 - \rs/r)\,\dot{t} = \mbox{constant}$.
The first of these invariants is the relativistic analogue to the Newtonian equation for the conservation of angular momentum per unit mass. The radial equation is most easily obtained by dividing the line element Eq.~(\ref{SchwarzschildMetric}) by $\rmd\tau^2$, resulting in
\begin{equation}
c^2 = ( 1 - \rs/r ) c^2 \dot{t}^2 - ( 1 - \rs/r )^{-1} \dot{r}^2 - r^2 \dot{\varphi}^2. \label{RadialEquation0}
\end{equation}
In terms of the invariants of motion, and with the definition $E \equiv \tfrac{1}{2}( k^2 - 1 ) c^2$, the radial equation Eq.~(\ref{RadialEquation0}) may be expressed as
\begin{equation}
E = \frac{1}{2} \dot{r}^2 - \frac{GM}{r} + \frac{\ell^2}{2 r^2} - \frac{\ell^2 \rs}{2 r^3}. \label{eq_rel_energy}
\end{equation}
Time is eliminated by the replacement
\begin{equation}
\dot{r} = - \ell \frac{\rmd}{\rmd\varphi} \frac{1}{r},
\end{equation}
resulting in
\begin{equation}
E = \frac{\ell^2}{2} \left( \frac{\rmd}{\rmd\varphi} \frac{1}{r} \right)^{\!2} - \frac{GM}{r} + \frac{\ell^2}{2 r^2} - \frac{\ell^2 \rs}{2 r^3}. \label{RadialEquation1}
\end{equation}
Differentiating Eq.~(\ref{RadialEquation1}) with respect to $\varphi$ results in a nonlinear differential equation for $1/r(\varphi)$,
\begin{equation}
0 = \ell^2 \frac{\rmd^2}{\rmd \varphi^2} \frac{1}{r} - GM + \frac{\ell^2}{r} - \frac{3 \ell^2 \rs}{2 r^2}. \label{eq_node}
\end{equation}

We anticipate a solution of Eq.~(\ref{eq_node}) that is near Keplerian and introduce the radius of a circular orbit for a classical particle with the same angular momentum, $r_\text{c} \equiv \ell^2/GM $. The result is\begin{equation}
\frac{\rmd^2}{\rmd \varphi^2} \frac{r_\text{c}}{r} + \frac{r_\text{c}}{r} = 1 + 3 \epsilon \left( \frac{r_\text{c}}{r} \right)^{\!2}\!, \label{eq_rel_orbits}
\end{equation}
where $\epsilon \equiv ( GM/\ell c )^2 = \rs/2r_\text{c}$. The conic-sections of Newtonian mechanics \cite{marion,goldstein,smiths} are recovered by setting $\epsilon = 0$:
\begin{equation}
\frac{\rmd^2}{\rmd \varphi^2} \frac{r_\text{c}}{r} + \frac{r_\text{c}}{r} = 1, \label{eq_conic_sections}
\end{equation}
which implies that
\begin{equation}
\frac{r_\text{c}}{r} = 1 + e\cos{\varphi}, \label{eq_conic_sections_sol}
\end{equation}
where $e$ is the eccentricity.

\section{Keplerian Limit and Perihelic Precession}\label{Sec_KepLim}

Relativistic orbits near a spherically symmetric massive object, such as our Sun, are described by Eq.~(\ref{eq_rel_orbits}). The planets of our solar system are described by near circular orbits. Mercury has the largest eccentricity $(e \approx 0.2)$, and the next largest is that of Mars $(e \approx 0.09)$. If $\epsilon$ is taken to be a small relativistic correction to the near-circular orbits of Newtonian mechanics, it is convenient to make the change of variable $1/\sigma \equiv r_\text{c}/r - 1 \ll 1$. The last term on the right-hand-side of Eq.~(\ref{eq_rel_orbits}) may now be approximated as $(r_\text{c}/r )^2 \approx 1 + 2/\sigma$, resulting in a linear differential equation for $1/\sigma(\varphi)$:
\begin{equation}
\frac{1}{3\epsilon}\frac{\rmd^2}{\rmd \varphi^2} \frac{1}{\sigma} + \frac{1-6\epsilon}{3\epsilon}\frac{1}{\sigma} \approx 1. \label{eq_linearized}
\end{equation}
The additional change of variable $\alpha \equiv \varphi \sqrt{1 - 6\epsilon}$ results in the familiar form:
\begin{equation}
\frac{\rmd^2}{\rmd {\alpha}^2} \frac{\sigma_\text{c}}{\sigma} + \frac{\sigma_\text{c}}{\sigma} \approx 1, \label{eq_fam_form}
\end{equation}
where $\sigma_\text{c} \equiv (1 - 6\epsilon)/3\epsilon$. The solution is similar to that of Eq.~(\ref{eq_conic_sections}):
\begin{equation}
\frac{\sigma_\text{c}}{\sigma} \approx 1 + A\cos{\alpha},\label{eq_eoo_sigma}
\end{equation}
where $A$ is an arbitrary constant of integration. In terms of the original coordinates, Eq.~(\ref{eq_eoo_sigma}) becomes
\begin{equation}
\frac{\tilde{r}_\text{c}}{r} \approx 1 + \tilde{e} \cos{\tilde{\kappa}\varphi}, \label{eq_S_eoo}
\end{equation}
where
\begin{align}
\tilde{r}_\text{c} &\equiv r_\text{c}\frac{1-6\epsilon}{1 - 3\epsilon}, \label{eq_S_coeff_r0}\\
\tilde{e} &\equiv \frac{3\epsilon A}{1 - 3\epsilon}, \label{eq_S_coeff_e0}\\
\tilde{\kappa} &\equiv (1 - 6\epsilon)^{\frac{1}{2}}. \label{eq_S_coeff_phi0}
\end{align}

According to the correspondence principle, Kepler's orbits, Eq.~(\ref{eq_conic_sections_sol}), must be recovered in the limit $\epsilon\rightarrow 0$, so that $3\epsilon A\equiv e$ is the eccentricity of Newtonian mechanics. To first order in $\epsilon$, Eqs.~(\ref{eq_S_coeff_r0})--(\ref{eq_S_coeff_phi0}) become
\begin{align}
\tilde{r}_\text{c} &\approx r_\text{c} (1 - 3\epsilon), \label{eq_S_coeff_r}\\
\tilde{e} &\approx e (1 + 3\epsilon), \label{eq_S_coeff_e} \\
\tilde{\kappa} &\approx 1 - 3\epsilon, \label{eq_S_coeff_phi}
\end{align}
and the relativistic orbits in this limit may be expressed concisely as
\begin{equation}
\frac{r_\text{c}(1-3\epsilon)}{r} \approx 1 + e(1 + 3 \epsilon)\cos{(1 - 3 \epsilon)\varphi}. \label{eq_S_eoo_concise}
\end{equation}
Equations~(\ref{eq_S_coeff_r})~and~(\ref{eq_S_coeff_e}) are the relativistic radius of circular orbit and relativistic eccentricity, respectively, and are discussed in more detail in Sec.~\ref{Sec_Discussion}.

\begin{figure}[ht]
\centering\includegraphics[width=\columnwidth]{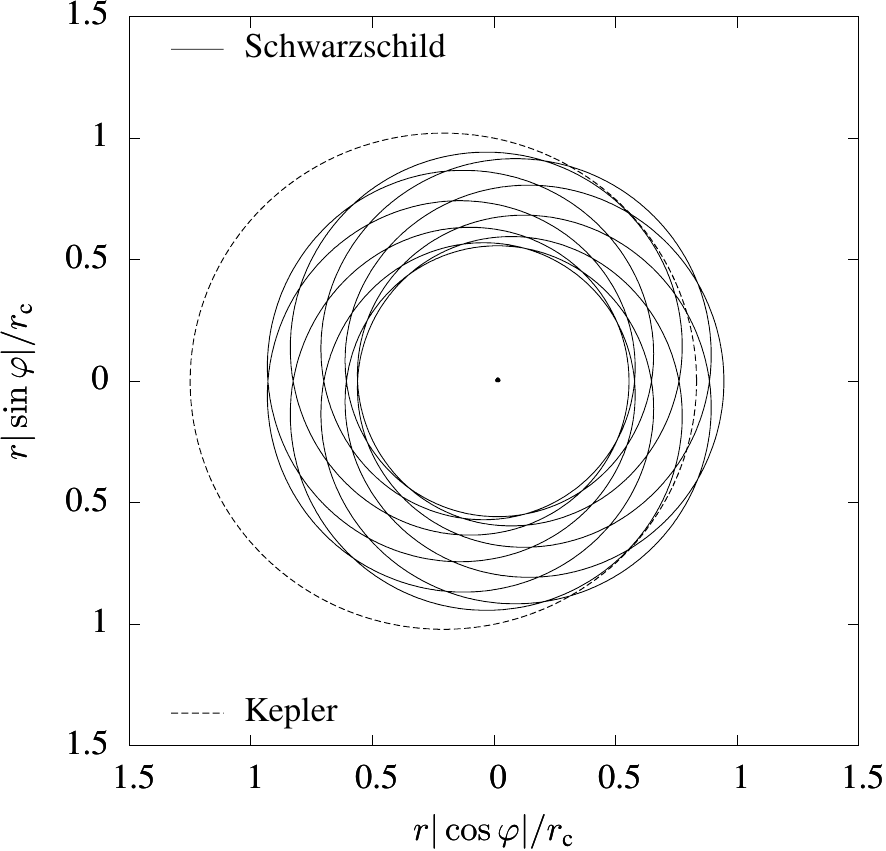}
\caption{\label{fig_precess} Relativistic orbit in a Keplerian limit (solid line) as described by Eq.~(\ref{eq_S_eoo_concise}), compared to a corresponding Keplerian orbit (dashed line), Eq.~(\ref{eq_conic_sections_sol}). The precession of the perihelion is one characteristic due to general relativity, and is illustrated here for $0\leq\varphi\leq 20\pi$. The eccentricity is chosen to be $e=0.2$ for both the relativistic and Keplerian orbits. This characteristic is exaggerated by the choice of relativistic correction parameter $(\epsilon=0.1)$ for purposes of illustration. Precession is present for smaller (non-zero) reasonably chosen values of $\epsilon$ as well.}
\end{figure}

The approximate orbit equation in Eq.~(\ref{eq_S_eoo_concise}) predicts a shift in the perihelion through an angle
\begin{equation}
\Delta\varphi \equiv 2\pi (\tilde{\kappa}^{-1} - 1) \approx 2\pi (3\epsilon) \label{eq_S_def_precess}
\end{equation}
per revolution. This first-order prediction is in agreement with well known results \cite{MTW,weinberg,rindler,PleKra,OR,dinverno,hartle,carroll,HobEfsLas,wald,ovanesyan,stump,nobili,magnan,deliseo,brillgoel,dean,ashby}, and agrees with the observed precession of perihelia of the inner planets \cite{MTW,weinberg,rindler,PleKra,OR,dinverno,HobEfsLas,stump,ovanesyan,hartle2,stewart,BroCle,sigismondi}. (Precession due to general relativity is illustrated in Fig.~\ref{fig_precess}.) As with other first-order calculations, this result may be compared to observations, assuming that the relativistic and Keplerian angular momenta are approximately equal \cite{MTW,weinberg,rindler,PleKra,OR,dinverno,hartle,carroll,HobEfsLas,ovanesyan}. For a Keplerian orbit \cite{marion,goldstein,smiths} $\ell^2 = GMa(1-e^2)$, where $G = 6.670\times 10^{-11}$\,m$^3\,$kg$^{-1}$\,s$^{-2}$, $M = 1.989\times 10^{30}$\,kg is the mass of the Sun, and $a$ and $e$ are the semi-major axis and eccentricity of the orbit, respectively. Therefore, the relativistic correction defined after Eq.~(\ref{eq_rel_orbits}),
\begin{equation}
3\epsilon \approx \frac{3GM}{c^2 a (1 - e^2)},
\end{equation}
is largest for planets closest to the Sun and for planets with very eccentric orbits. For Mercury \cite{verb} $a = 5.79 \times 10^{10}$\,m and $e = 0.2056$, so that $3\epsilon \approx 7.97\times 10^{-8}$. (The speed of light is taken to be $c^2 = 8.987554\times 10^{16}\,\text{m}^2/\text{s}^2$.) According to Eq.~(\ref{eq_S_def_precess}), Mercury precesses through an angle
\begin{equation}
\Delta\varphi \approx \frac{6\pi GM}{c^2 a (1 - e^2)} = 5.02 \times 10^{-7}\,\text{rad}
\end{equation}
per revolution. This angle is very small and is usually expressed cumulatively in arc seconds per century. The orbital period of Mercury is 0.24085 terrestrial years, so that
\begin{subequations}
\begin{align}
\Delta\Phi &\equiv \dfrac{100\,\text{yr}}{0.24085\,\text{yr}} \times \dfrac{360 \times 60 \times 60}{2\pi} \times \Delta\varphi\\
&\approx 43.0\,\text{arcsec/century}.
\end{align}
\end{subequations}
Historically, this contribution to the precession of perihelion of Mercury's orbit precisely accounted for the observed discrepancy, serving as the first triumph of the general theory of relativity \cite{einstein0,einstein1,einstein2,einstein3}.

\section{Discussion}\label{Sec_Discussion}

The approximate orbit equation in Eq.~(\ref{eq_S_eoo_concise}) provides small corrections to Kepler's orbits due to general relativity. [Compare Eqs.~(\ref{eq_S_eoo_concise})~and~(\ref{eq_conic_sections_sol}).] A systematic verification may be carried out by substituting Eq.~(\ref{eq_S_eoo_concise}) into Eq.~(\ref{eq_rel_orbits}), keeping terms of orders $e$, $\epsilon$, and $e\epsilon$ only. The justification for discarding the term nonlinear in the eccentricity is the correspondence principle. Arguments concerning which terms to discard based only on direct comparisons of relative magnitudes of higher-order and lower-order terms lead to contradictions (as may be verified by the reader). Instead, the domain of validity is expressed by subjecting the solution Eq.~(\ref{eq_S_eoo_concise}) to the condition
\begin{equation}
\frac{r_\text{c}}{r} - 1 \ll 1 \label{eq_DOV}
\end{equation}
for the smallest value of $r$. Evaluating the equation of the orbit Eq.~(\ref{eq_S_eoo_concise}) at the perihelion $(r_\text{p})$ results in
\begin{equation}
\frac{r_\text{c}}{r_\text{p}} = \frac{1 + e(1 + 3\epsilon)}{1 - 3\epsilon}.
\end{equation}
The substitution of this result into Eq.~(\ref{eq_DOV}) results in the domain of validity:
\begin{equation}
e(1 + 3\epsilon) + 2(3\epsilon) \ll 1. \label{eq_S_1st_valid}
\end{equation}
Therefore, the relativistic eccentricity $\tilde{e} = e(1+3\epsilon)\ll 1$, and Eq.~(\ref{eq_S_eoo_concise}) is limited to describing relativistic corrections to near-circular (Keplerian) orbits. Also, the relativistic correction $2(3\epsilon)\ll 1$, and thus the equation of orbit Eq.~(\ref{eq_S_eoo_concise}) is valid only for small relativistic corrections.

\begin{figure}[ht]
\centering\includegraphics[width=\columnwidth]{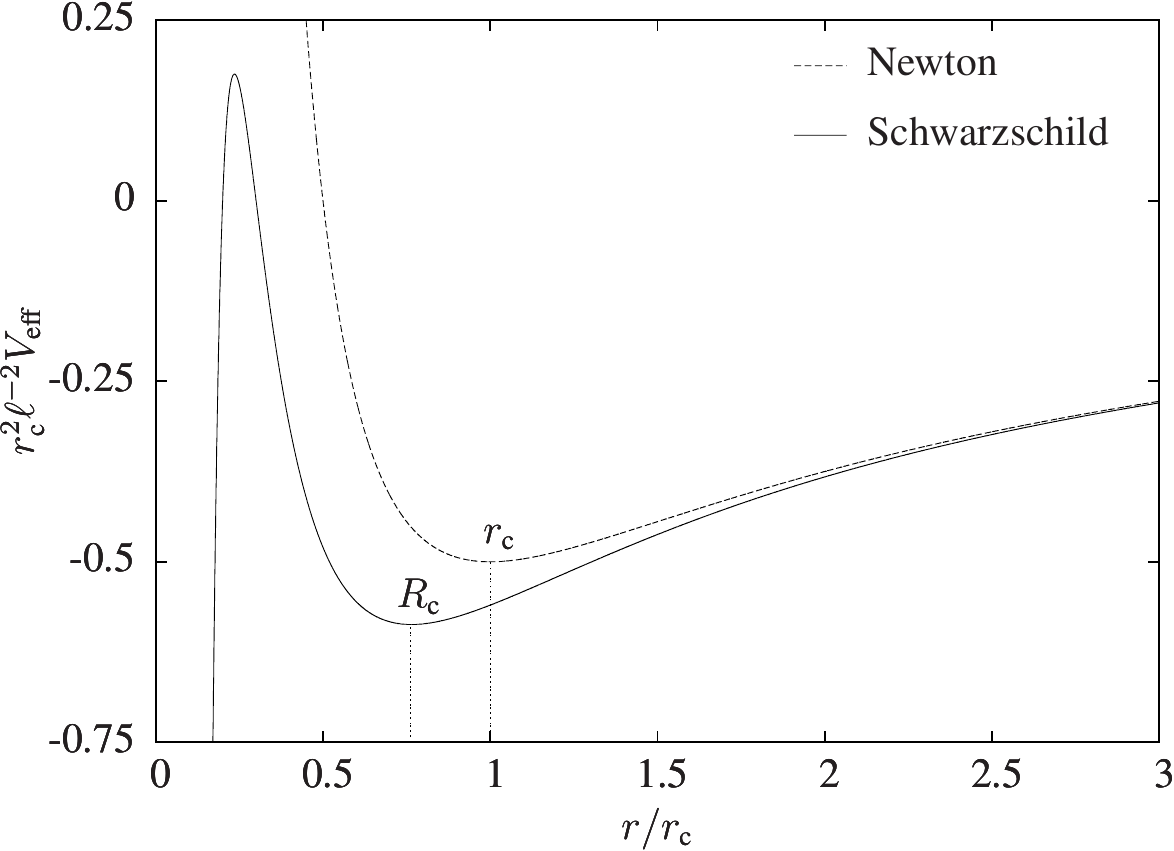}
\caption{\label{fig_S_eff_pot} The effective potential in Eq.~(\ref{eq_Schwarz_Eff_Pot}) derived from the Schwarzschild geometry (solid line) compared to that derived from Newtonian mechanics (dashed line). The vertical dotted lines identify the radii of circular orbits, $R_\text{c}$ and $r_\text{c}$, as calculated using the Schwarzschild geometry and Newtonian mechanics, respectively. The Schwarzschild geometry predicts a smaller radius of the circular orbit than that predicted by Newtonian mechanics [Eq.~(\ref{eq_S_rc_eff_pot})]. The value $\epsilon=0.08$ is chosen for purposes of illustration.}
\end{figure}

Equation~(\ref{eq_S_eoo_concise}) is also consistent with a reduced radius of the circular orbit, as derived from the Schwarzschild effective potential:
\begin{equation}
V_\text{eff} \equiv - \frac{GM}{r} + \frac{\ell^2}{2 r^2} - \frac{\ell^2 \rs}{2 r^3}. \label{eq_Schwarz_Eff_Pot}
\end{equation}
This effective potential is customarily defined \cite{rindler,OR,hartle,carroll,HobEfsLas,wald,ovanesyan,MTW2} after making the observation that Eq.~(\ref{eq_rel_energy}) reduces to the Newtonian result as $\rs \rightarrow 0$ $(c \rightarrow \infty)$. [The Schwarzschild effective potential in Eq.~(\ref{eq_Schwarz_Eff_Pot}) is compared to that derived from Newtonian mechanics in Fig.~\ref{fig_S_eff_pot}.] If we minimize $V_\text{eff}$ with respect to $r$, we obtain the radius of the circular orbit,
\begin{equation}
R_\text{c} = \frac{1}{2} r_\text{c} + \frac{1}{2} r_\text{c} \sqrt{1 - 12 \epsilon} \approx r_\text{c} (1 - 3\epsilon), \label{eq_S_rc_eff_pot}
\end{equation}
in agreement with Eq.~(\ref{eq_S_coeff_r}) to first order in $\epsilon$. An additional characteristic of relativistic orbits is that of increased eccentricity Eq.~(\ref{eq_S_coeff_e}).  Equation~(\ref{eq_S_eoo_concise}) predicts that relativistic orbits will have increased eccentricity compared to Keplerian orbits. The correction to the eccentricity is of the same order as that for the relativistic contribution to perihelic precession and the reduced radius of the circular orbit.

\section{Conclusion}\label{Sec_Conclusion}

The relativistic central-mass problem has been cast into finding a solution that is not very different from a Newtonian circular orbit, the result of which is a Keplerian orbit with small relativistic corrections. The resulting first-order relativistic contribution to the perihelic precession in Eq.~(\ref{eq_S_def_precess}) agrees with established calculations. This effect is one of several relativistic effects which arise as first-order corrections to the familiar orbit equation describing Keplerian orbits in Eq.~(\ref{eq_conic_sections_sol}). The approximate orbit equation is limited to describing small relativistic corrections to near-circular Keplerian orbits, as expressed by the domain of validity in Eq.~(\ref{eq_S_1st_valid}). Comparisons with observations are made with the standard assumption that the relativistic and Keplerian angular momenta are approximately equal.

Compared to the usual perturbative approach, this derivation demonstrates several deliberate steps toward a Keplerian limit prior to solving the nonlinear differential equation (\ref{eq_node}) arising from the Schwarzschild line element. The linearization procedure preceding Eq.~(\ref{eq_linearized}) and appeal to the correspondence principle [to identify the Newtonian eccentricity in Eq.~(\ref{eq_S_coeff_e0})] makes for a very physical approach toward the boundary between relativistic and Keplerian orbits. The resulting approximate orbit equation in Eq.~(\ref{eq_S_eoo_concise}) lends itself to easy comparison with the familiar Keplerian orbits in Eq.~(\ref{eq_conic_sections_sol}) and displays several relativistic effects. This simple solution arises as a result of pursuing a narrower problem from the beginning, and explicitly expressing the domain of validity in Eq.~(\ref{eq_S_1st_valid}). This derivation is approachable to undergraduate students, and the resulting orbit equation provides qualitative and quantitative understanding of the corrections to Keplerian orbits due to general relativity.

\begin{acknowledgments}
We would like to thank Neil Ashby, Shane Burns, Dick Hilt, and Patricia Purdue for their valuable conversations, advice, and corrections. We are most grateful to the reviewers for their careful attention to our manuscript, resulting in a presentation that is much improved in both content and style.
\end{acknowledgments}

\begin{thebibliography}{99}
\bibitem{verb} The algebra is included here for completeness, and is taken almost verbatim from the very clear presentations of Refs.~\onlinecite{PleKra} and \onlinecite{carroll}.
\end{thebibliography}
\end{document}